\documentstyle[sprocl,epsfig]{article}

\begin{document}

\def\Journal#1&#2&#3(#4){#1{\bf #2} (#4) #3}

\def\AP{{\em Ann. Phys. }}
\def\NIM{{\em Nucl. Inst. and Meth. }}
\def\JPG{{\em J. Phys. }{\bf G}}
\def\NIMA{{\em Nucl. Inst. and Meth. }{\bf A}}
\def\NCA{{\em Nuovo. Cim. }{\bf A}}
\def\JETFL{{\em JETF Lett. }}
\def\NPA{{\em Nucl. Phys. }{\bf A}}
\def\NPB{{\em Nucl. Phys. }{\bf B}}
\def\PLB{{\em Phys. Lett. }{\bf B}}
\def\PL{{\em Phys. Lett. }}
\def\PRL{{\em Phys. Rev. Lett. }}
\def\TMP{{\em Theor. Math. Phys. }}
\def\PR{{\em Phys. Rep. }}	
\def\PRD{{\em Phys. Rev. }{\bf D}}
\def\PREV{{\em Phys. Rev. }}
\def\PRC{{\em Phys. Rep. }{\bf C}}
\def\ZPA{{\em Z. Phys. }{\bf A}}
\def\ZPC{{\em Z. Phys. }{\bf C}}
\def\IJMF{{\em Int. J. Mod. Phys. }}

\def\earc#1&#2(#3){#1{\bf #2} (#3).}

\def\newblock{~}

\def\etal   {{\it et al.}}
\def\etaln  {{\it et al}}

\def\gevt   {{\rm GeV^2}}

\def\piz    {\pi^0}
\def\etaz   {\eta}
\def\etap   {\eta^{\prime}}
\def\ra     {\rightarrow}
\def\gam    {\gamma}
\def\pip    {\pi^+}
\def\pim    {\pi^-}
\def\rhoz   {\rho^0}
\def\sig    {\sigma}
\def\qsq    {Q^2}

\def\qsqp   {{Q^\prime}^2}

\def\gaga   {\gamma\gamma}
\def\gsg    {\gamma^*\gamma}
\def\mgsg   {\gamma^*\gamma\rightarrow{\cal R}}
\def\mgsgs  {\gamma^*\gamma^*\rightarrow{\cal R}}
\def\gsgs   {\gamma^*\gamma^*}
\def\qq     {q\tilde{Q}}

\def\frg    {{\cal F}_{\gamma^*\gamma{\cal R}}}
\def\frgs   {{\cal F}_{\gamma^*\gamma^*{\cal R}}}
\def\fpizg  {{\cal F}_{\gamma^*\gamma\pi^0}}
\def\fpizgs {{\cal F}_{\gamma^*\gamma^*\pi^0}}
\def\fetazg {{\cal F}_{\gamma^*\gamma\eta}}
\def\fetapg {{\cal F}_{\gamma^*\gamma\eta^{\prime}}}

\title{MEASUREMENTS OF THE MESON-PHOTON TRANSITION
FORM FACTORS OF LIGHT PSEUDOSCALAR MESONS 
AT LARGE MOMENTUM TRANSFER}

\author{V. Savinov (representing the CLEO Collaboration) \\ (Contribution to the conference Photon97, Egmond aan Zee (1997))}

\address{Stanford Linear Accelerator Center, \\ MS61, P.O. Box 4349, Stanford, CA, 94309, USA \\
(e-mail: savinov@lns62.lns.cornell.edu)}

\maketitle\abstracts{
Using the CLEO~II detector, we 
have measured the form factors associated with the 
electromagnetic transitions 
\mbox{$\gamma^*\gamma$ $\ra$ meson}. 
We have measured these form factors 
in the momentum transfer ranges 
from 1.5 to 
$9$, $20$, and $30$ $\gevt$ 
for $\piz$, $\etaz$, and $\etap$, 
respectively. 
}

\section{Introduction}

Production of even $C$-parity hadronic matter in $e^+e^-$ scattering provides 
a unique opportunity to study the properties of strong interactions. 
To leading order in quantum electrodynamics (QED) these processes 
are described as the interaction between two photons 
emitted by the scattered electrons. 
Although in $e^+e^-$ scattering the probe and the target are both 
represented by photons that are carriers of the electromagnetic force, 
these space-like photons can produce a pair of quarks 
that interact strongly and are observed in the form of hadrons. 
Therefore, by measuring the four-momenta of the 
scattered electrons we can study the 
dynamics of strong interactions. 
The quantities of interest in these studies 
are the form factors associated with the 
transitions between the photons and the hadrons. 

In this paper we briefly describe the final results 
of our measurements~\cite{SAVINOV:thesis,PRD-CLNS} of 
the differential cross sections for the production of a single pseudoscalar 
meson in $e^+e^-$ scattering: 
$e^+e^- \ra e^+e^-{\cal R}$, 
where ${\cal R}$ is a $\piz$, $\etaz$ or $\etap$. 
We measure these cross sections in a ``single-tagged'' experimental mode 
where one of the scattered electrons is detected (``tagged''), 
while the other electron is scattered at a very small angle and 
therefore remains undetected (``{\em un}tagged''). 
The mesons produced in $e^+e^-$ scattering are observed 
through their decays to various fully reconstructed final states. 
The tagged electron 
emits a highly off-shell photon ($\gamma^*$), whereas the untagged 
electron emits 
a nearly on-shell photon ($\gamma$). We measure the dependence of the meson 
production rate on the squared momentum transfer $Q^2$ carried by 
the highly off-shell photon. 
This momentum transfer is determined by energy-momentum conservation as applied to 
the tag: 
$Q^2 \equiv 
-( p_b - p_t)^2 = 
2 E_b E_t (1-\cos\theta_t)$,
where $p_b$ and $p_t$ are the four-momenta 
of the incident 
beam-energy electron and the tag, 
$E_b$ and $E_t$ are corresponding energies, 
and $\theta_t$ is the scattering angle. 
From the measurements of the differential rates 
$d\sigma(e^+e^- \ra e^+e^-{\cal R})/d\qsq$ 
we obtain the transition form factors $\frg$ 
that describe the effect of the strong interaction in the 
$\mgsg$ transition amplitudes. 

To relate the differential cross sections to 
the transition form factors we employ the theoretical 
framework developed by V.M. Budnev \etal ~\cite{BGMS:75} (BGMS formalism). 
In BGMS the process $e^+e^- \ra e^+e^-{\cal R}$ is divided 
into two parts: $e^+e^- \ra e^+e^-\gamma^*\gamma$ and 
$\gamma^*\gamma \ra {\cal R}$. 
The first part is completely calculable in QED and the second part is 
defined in terms of the transition form factors $\frg(\qsq)$. 
In the case of pseudoscalar mesons there is only one form factor. 
At zero momentum transfer this form factor is expressed as: 
\begin{equation}
|\frg(0)|^2 =\,
\frac{1}{(4 \pi \alpha)^2} \,
\frac{64 \pi \Gamma(\cal R \ra \gaga)}{M_{\cal R}^3} \,
, 
\label{EQ:1}
\end{equation} 		
where $\alpha$ is the QED coupling constant, 
$M_{\cal R}$ is the mass 
and 
$\Gamma(\cal R \ra \gamma\gamma)$ 
is 
the two-photon partial width of the meson ${\cal R}$. 
The transition form factors cannot be calculated directly from 
Quantum Chromodynamics (QCD). 
However, these form factors have been 
estimated~\cite{KROLL:WUB9417,KROLL:96,GUANG:96,RR:9603408} 
using theoretical methods based on 
the perturbative QCD (PQCD)~\cite{BL:80} 
and 
the sum rules~\cite{AV:sum_rules}. 

Brodsky and Lepage employed PQCD to find 
the asymptotic behavior of the $\mgsg$ transition 
form factors in the limit $\qsq\ra\infty$~\cite{BL:81}: 
\begin{equation} 
\lim_{\qsq \ra \infty} \qsq \frg(\qsq) = 2 f_{\cal R}, 
\label{EQ:2} 
\end{equation} 
where $f_{\cal R}$ is the meson decay constant. 
In addition, it has been predicted  
that in this limit any mesonic wave function 
evolves to the asymptotic wave function of unique shape~\cite{BL:80,ER:79,CZ:84}. 

While PQCD predicts the form factors of the $\mgsg$ transitions at large momentum transfer, 
the behavior of these form factors in the limit $\qsq\ra0$ 
can be determined from the axial anomaly~\cite{ANOMALY:axial1,ANOMALY:axial2} 
in the chiral limit of QCD. 
For $\piz$ and $\etaz$ the axial anomaly yields~\cite{BL:81}: 
\begin{equation}
\lim_{\qsq \ra 0} \frg(\qsq) = \frac{1}{4 \pi^2 f_{\cal R}}, 
\label{EQ:3}
\end{equation} 
to leading order in $m_u^2/M_{\cal R}^2$ and $m_d^2/M_{\cal R}^2$ where 
$m_u$ and $m_d$ are the masses of the $u$ and $d$ quarks. 
This prediction does not hold with the same precision 
for $\etap$ due to the larger value of the $s$-quark mass. 
In addition, even if the $s$-quark mass were small, 
this prediction might be broken for $\etap$ 
because this particle is an unlikely candidate for the Goldstone boson~\cite{RYDER:U1,COLEMAN:U1}. 

Finally, to approximate the soft non-perturbative region of $\qsq$ 
a simple interpolation between $\qsq \ra 0$ and $\qsq \ra \infty$ limits has been 
proposed~\cite{BL:81}: 
\begin{equation}
\frg(\qsq) \sim \frac{1}{4 \pi^2 f_{\cal R}} \frac{1}{1+(\qsq/8 \pi^2 f_{\cal R}^2)}. 
\label{EQ:4}
\end{equation} 

We have measured the transition form factors $\frg$ 
in the space-like regions of the momentum transfer 
between $1.5$ and $9$ $\gevt$ for $\piz$, 
$1.5$ and $20$ $\gevt$ for $\etaz$, 
and 
$1.5$ and $30$ $\gevt$ for $\eta^{\prime}$. 
We report the measurements of the transition form factors 
of $\piz$, $\etaz$, and $\etap$ using the decays: 
$\piz \ra \gamma\gamma$,
$\etaz \ra \gamma\gamma$,
$\etaz \ra \piz\piz\piz \ra 6\gamma$,
$\etaz \ra \pip\pim\piz \ra \pip\pim2\gamma$,
$\etap \ra \rhoz\gamma \ra \pip\pim\gamma$,
$\etap \ra \pip\pim\etaz \ra \pip\pim2\gamma$, 
$\etap \ra \piz\piz\etaz \ra 6\gamma$, 
$\etap \ra \pip\pim\etaz \ra 2\pip2\pim2\gamma$, 
$\etap \ra \piz\piz\etaz \ra 5\piz \ra 10\gamma$, 
$\etap \ra \piz\piz\etaz \ra 3\piz\pip\pim \ra \pip\pim6\gamma$, 
and 
$\etap \ra \pip\pim\etaz \ra \pip\pim3\piz \ra \pip\pim6\gamma$. 
We have analyzed the last two decay chains of $\etap$ together since 
they are observed in the same final state $\pip\pim6\gamma$. 

The data sample employed in our analysis 
corresponds to an integrated $e^+e^-$ luminosity 
of $2.88 \pm 0.03$ fb$^{-1}$ collected 
at $e^+e^-$ center-of-mass energy around 10.6 GeV 
with the CLEO-II detector~\cite{CLEO-II:detector,CLEO-II:trigger} at CESR. 

\section{Analysis Procedure}

To measure the products of the differential cross sections 
and branching fractions for each decay chain 
we use the following analysis procedure. 
Data events that pass all selection criteria \cite{PRD-CLNS} 
are used 
to form the $\qsq$ distribution where the value of $\qsq$ for 
each event is estimated from energy-momentum conservation for this event 
(the experimental method we use to estimate the value of $\qsq$ for each event 
is described in detail elsewhere~\cite{SAVINOV:thesis,PRD-CLNS}). 
Next we divide the event yields into $\qsq$ intervals. 
For each $\qsq$ interval 
we obtain the number of signal events in data 
from the fit to the invariant mass distribution. 
Then we estimate and subtract the (small) feed-down background~\cite{PRD-CLNS}. 
Finally we 
correct the background-subtracted number of signal events 
for the detection efficiency. 
The signal line shapes used in the fits 
and the detection efficiencies 
are determined 
from the detector simulation 
for each $\qsq$ interval. 
To extract the transition form factors we compare 
the measured and the predicted values of the cross sections. 
Namely, for each $\qsq$ interval, we measure the form factors 
$\frg^{data}(\tilde{Q}^2)$ from: 
$
|\frg^{data}(\tilde{Q}^2)|^2 = \sigma(data)/\sigma(MC) |\frg^{MC}(\tilde{Q}^2)|^2$,
where 
$\frg^{MC}(\tilde{Q}^2)$ is the approximation for the $\qsq$-dependent part 
of the form factor in Monte Carlo (MC) simulation, 
and $\sigma(data)$ and $\sigma(MC)$ are the cross sections for 
this $\qsq$ interval measured in data and predicted using 
numerical integration, respectively. 
The transition form factors are measured at $\tilde{Q}^2$ where 
the differential cross sections achieve their mean values 
according to the results of numerical integration. 

The $\qsq$-dependent part of the $\mgsgs$ transition form factors in our 
two-photon MC simulation program is approximated by: 
\begin{eqnarray}
|\frgs(Q^2)|^2 & = & \,
\frac{1}{(4 \pi \alpha)^2} \,
\frac{64 \pi \Gamma(\cal R \ra \gaga)}{M_{\cal R}^3} \,
\frac{1}{(1+Q^2/\Lambda_{\cal R}^2)^2} \,
,
\label{EQ:5}
\end{eqnarray} 
where the pole-mass parameter \mbox{$\Lambda_{\cal R}$ = 770 MeV}. 

\section{Results}
 
\begin{figure}[h]
\centerline{
	\psfig{figure=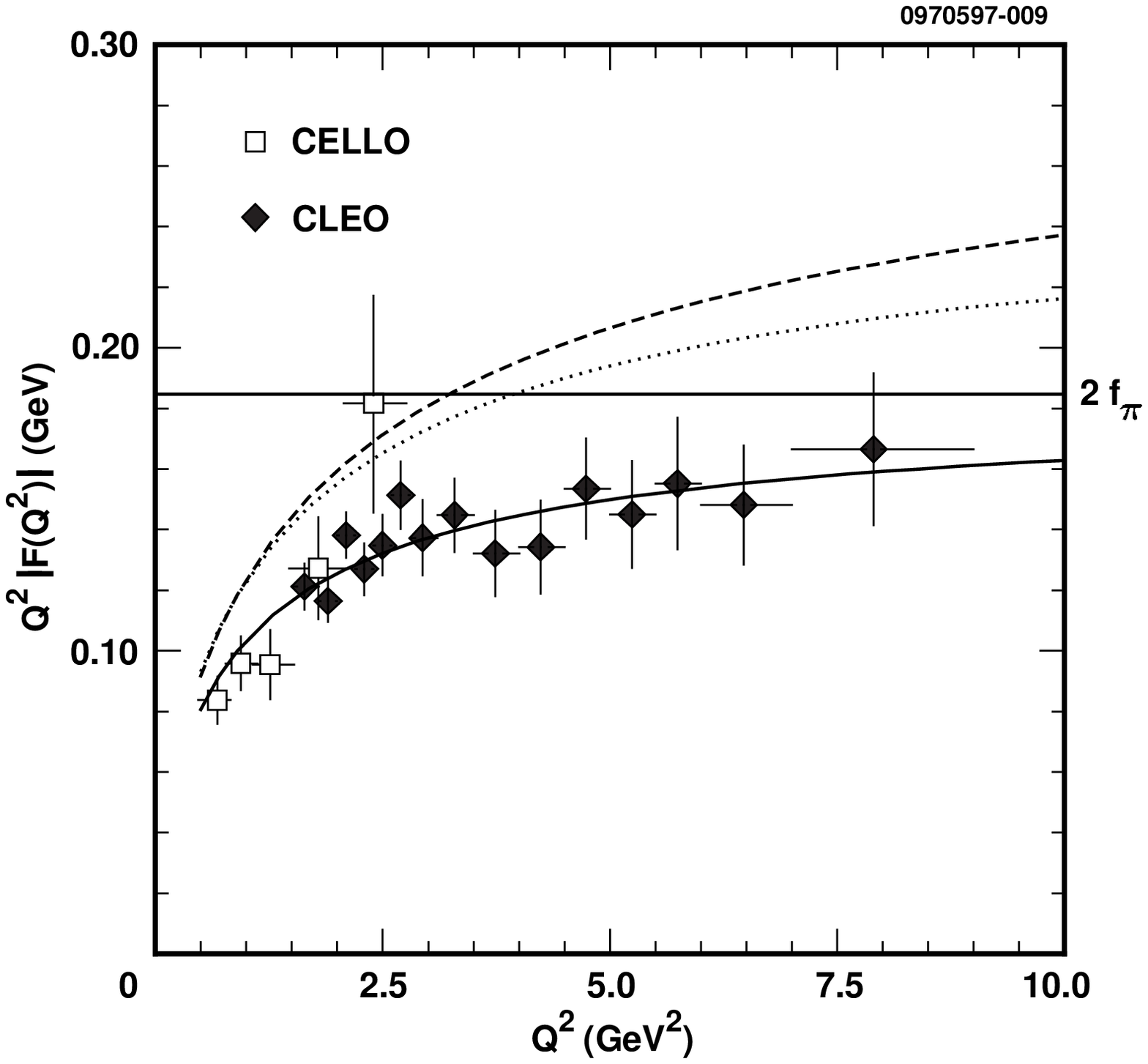,height=1.45in}
	\psfig{figure=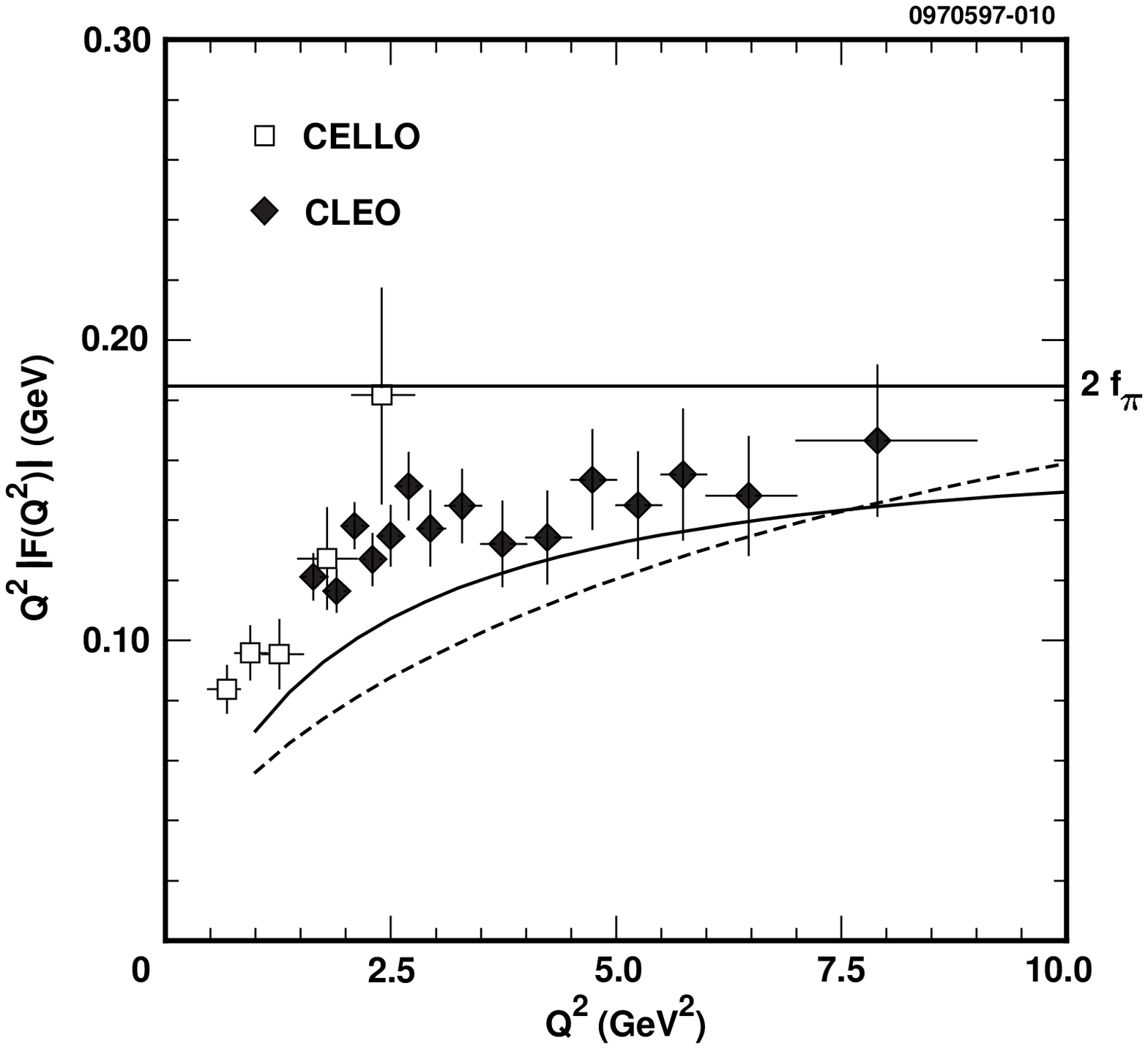,height=1.45in}
	\psfig{figure=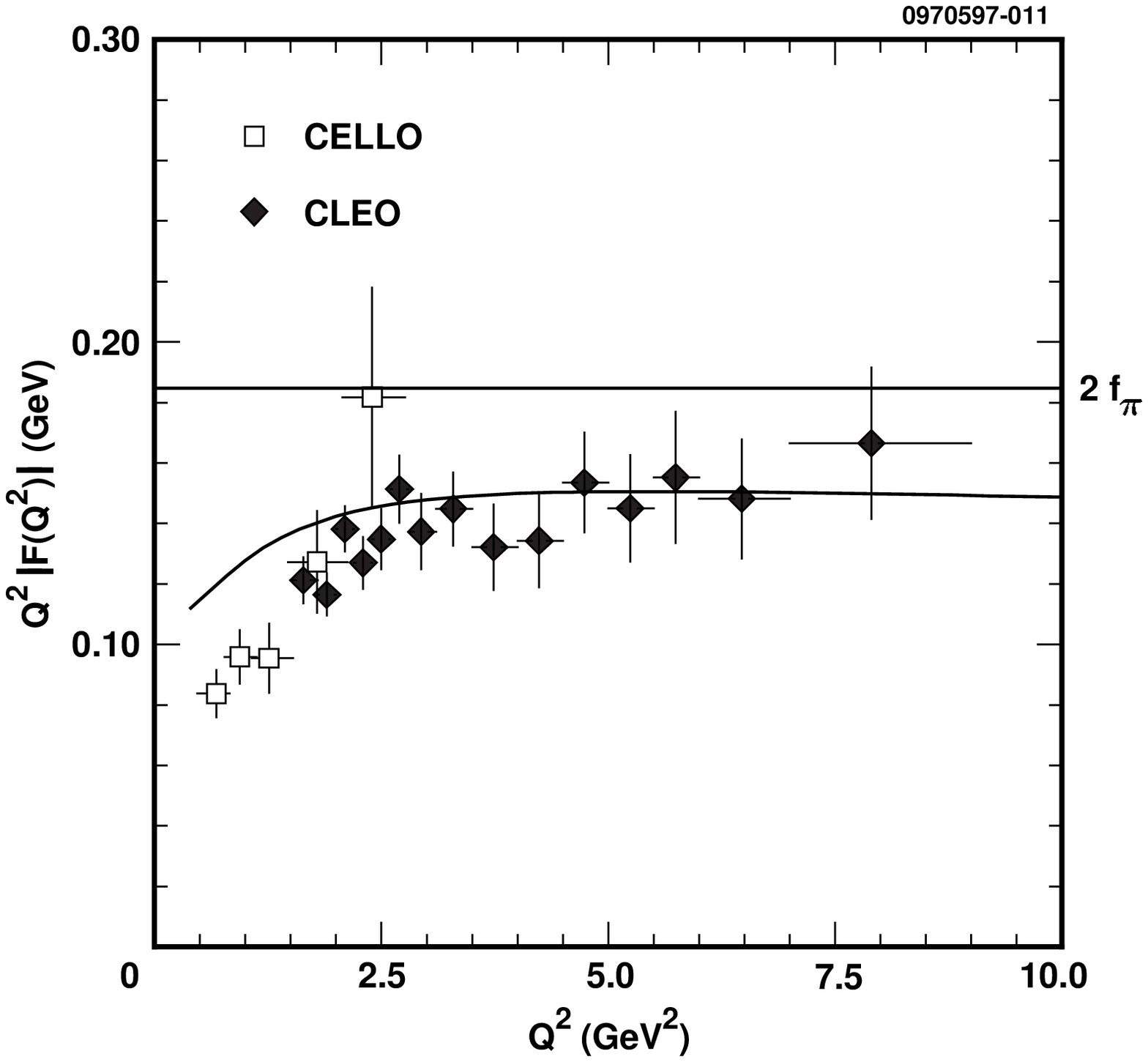,height=1.45in}
           }
\caption{
Comparison of the results (points) for $\piz$ with 
the theoretical predictions~\protect\cite{KROLL:96,GUANG:96,RR:9603408} 
(from left to right). 
Two figures on the left show the predictions with 
the asymptotic wave function~\protect\cite{BL:80,ER:79,CZ:84} (solid curves) 
and 
the CZ wave function~\protect\cite{CZ:84} (dashed curves). 
The dotted curve shows the prediction made with 
the CZ wave function when its QCD evolution is taken into account. 
}
\label{fig:ph97_fig_1}
\end{figure}

\begin{figure}[h]
\centerline{
	\psfig{figure=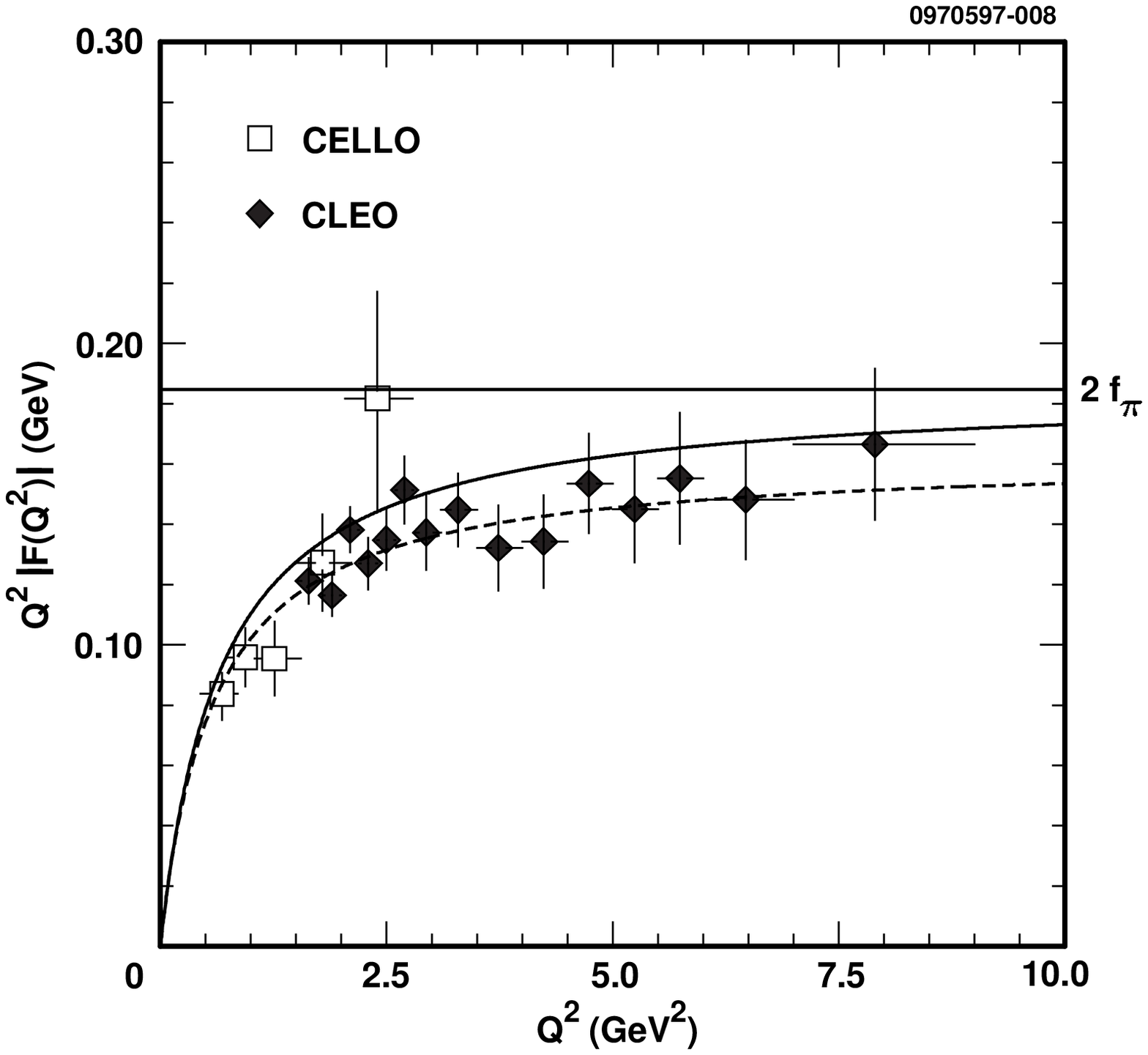,height=1.45in}
	\psfig{figure=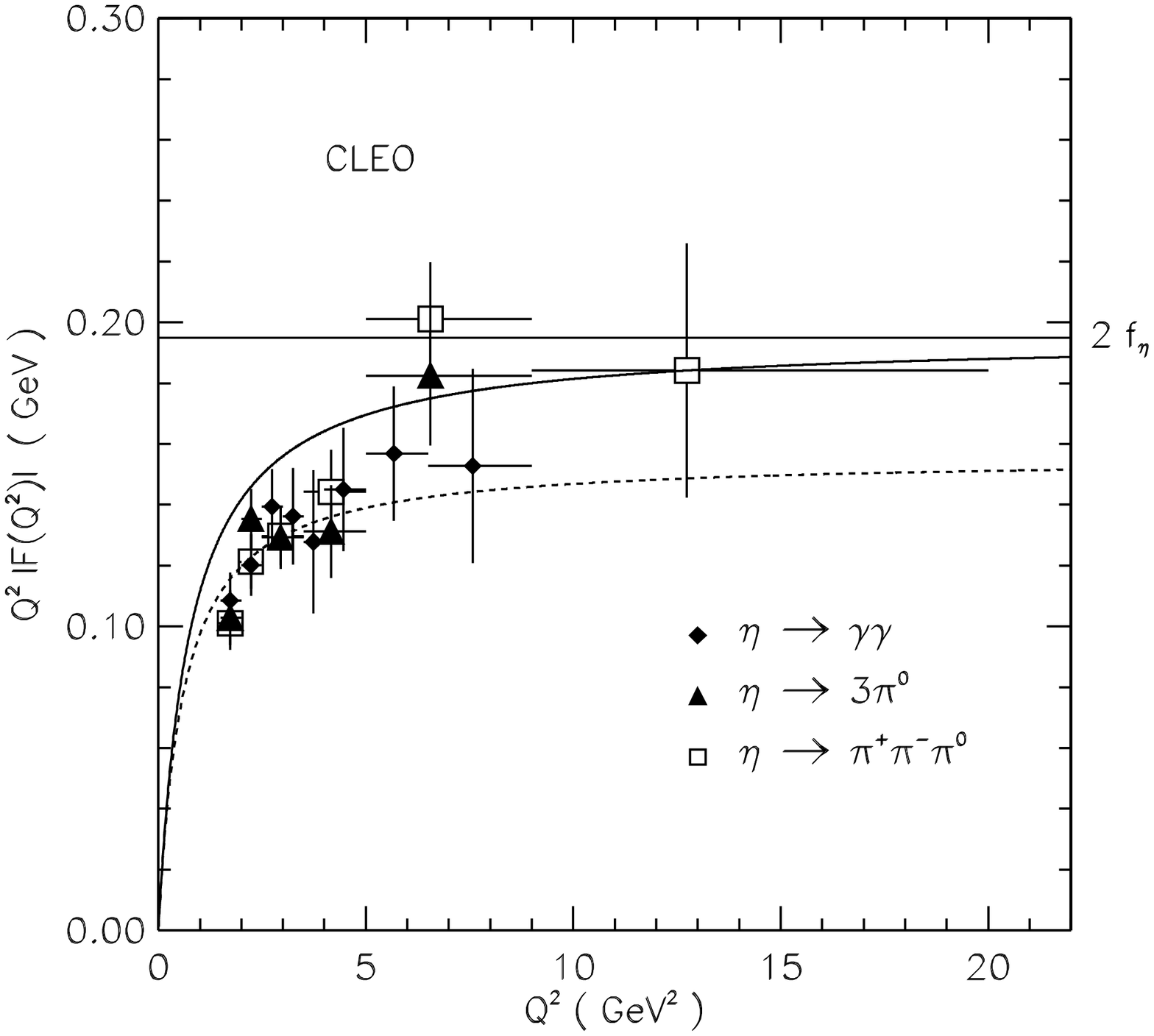,height=1.45in}
	\psfig{figure=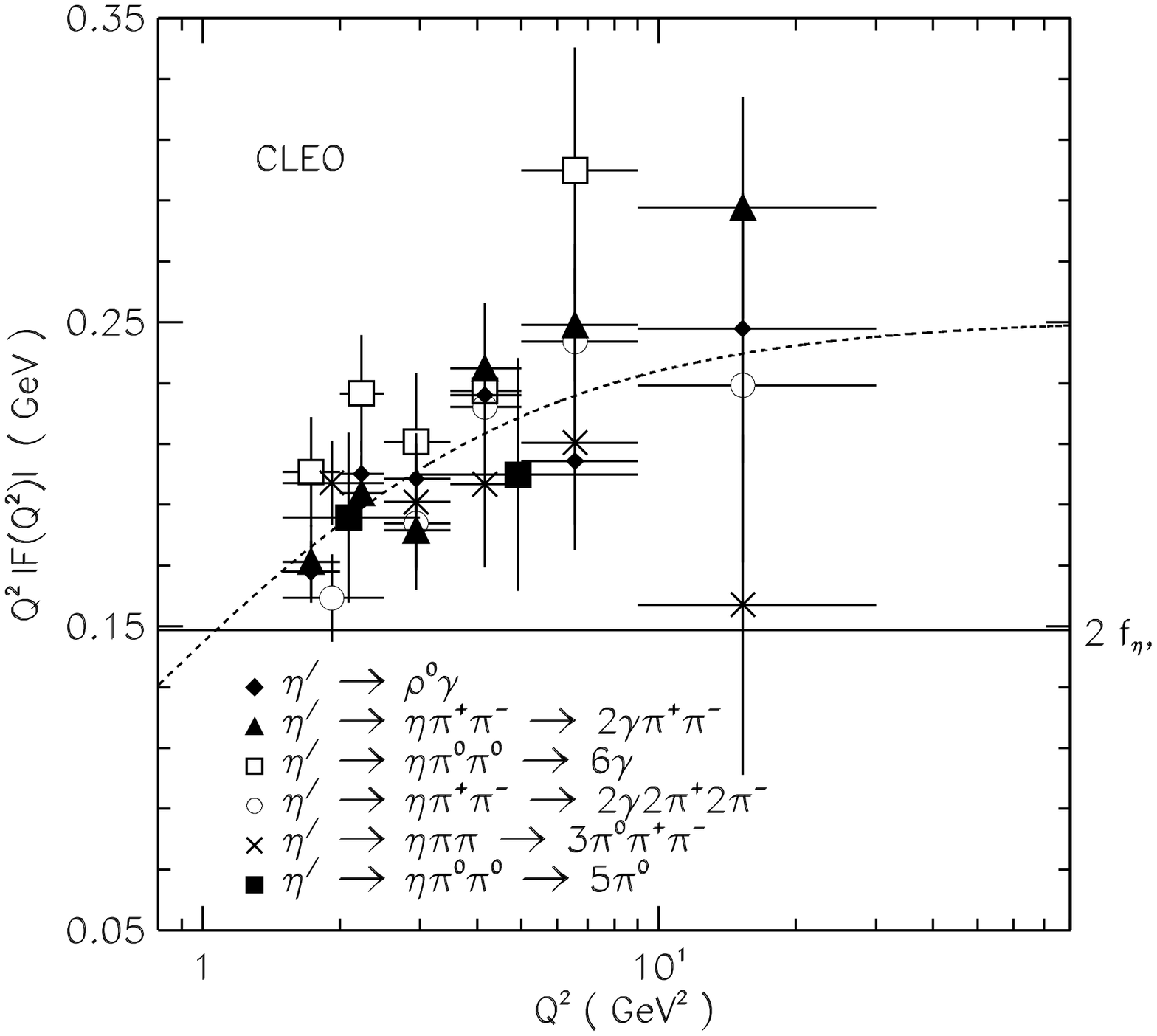,height=1.45in}
           }
\caption{
Pole-mass parameter fits (dashed lines) to CLEO results 
for (from left to right) $\piz$, $\etaz$, and $\etap$. 
The solid curves (where shown) are the interpolations given by Eqn.~\ref{EQ:4}. 
}
\label{fig:ph97_fig_2}
\end{figure}

In Figures~\ref{fig:ph97_fig_1}~and~\ref{fig:ph97_fig_2} 
we compare our results with the theoretical predictions. 
In these figures we also show the results of the CELLO 
experiment~\cite{CELLO:ff} and the asymptotic prediction 
of PQCD given by Eqn.~\ref{EQ:2}. 
For both experimental results the error bars represent 
the statistical errors only. 
To plot the results of the theoretical predictions we use 
their published analytical forms. 
To estimate the values of the meson decay constants $f_{\cal R}$ 
we use Eqns.~\ref{EQ:1}~and~\ref{EQ:3} 
and the tabulated two-photon partial widths of the studied mesons~\cite{PDG:96}. 

Finally, for each meson ${\cal R}$, where ${\cal R}$ is $\piz$, $\etaz$, or $\etap$, 
we derive the values of the pole-mass parameters $\Lambda_{\cal R}$ 
which we use to represent our results in a simple phenomenological form. 
For each meson we fit all our results for $|\frg(\qsq)|^2$ 
with the function given by Eqn.~\ref{EQ:5} and obtain the 
values of the pole-mass parameters $\Lambda_{\cal R}$ 
shown in Table~\ref{tab:final_res_pole}. 
In this table, for each measurement, the first error is statistical, 
the second error represents the systematic uncertainties of our measurement 
and the third error reflects the experimental 
error in the value of the two-photon partial width of the meson. 
The results of the fits are also shown in Figures~\ref{fig:ph97_fig_1}~and~\ref{fig:ph97_fig_2}. 

We use the measured values of the parameters $\Lambda_{\piz}$ and $\Lambda_{\etaz}$ 
to compare the soft non-perturbative properties of $\piz$ and $\etaz$. 
This is a legitimate comparison 
because 
the asymptotic prediction given by Eqn.~\ref{EQ:2} 
and 
the chiral limit given by Eqn.~\ref{EQ:3} 
are expected to hold for both $\piz$ and $\etaz$. 
From the comparison between the measured values of $\Lambda_{\piz}$ and $\Lambda_{\etaz}$ 
we conclude that the $\qsq$ shapes of the \mbox{$\gamma^*\gamma$ $\ra$ meson} 
transition form factors of $\piz$ and $\etaz$ are nearly identical, 
which strongly indicates the similarity between the wave functions of these mesons. 

The results of our measurements for the production of $\etap$ 
demonstrate that 
if this particle were a $q\bar{q}$ bound state 
and 
the QCD chiral limit given by Eqn.~\ref{EQ:3} held for this meson, 
the $\qsq$-dependence of the transition form factor of $\etap$ 
and consequently its wave function would be significantly 
different from these non-perturbative properties of either~$\piz$~or~$\etaz$. 

\section{Conclusions}

We have measured the form factors associated with 
the electromagnetic transitions \mbox{$\gamma^*\gamma$ $\ra$ meson} 
in the regions of momentum transfer from 1.5 to 
9, 20, and 30 $\gevt$ 
for $\piz$, $\etaz$, and $\etap$, respectively. 
These are the first measurements 
above 2.7 $\gevt$ for $\piz$ and above 7 $\gevt$ 
for $\etaz$ and $\etap$. 

Our measurement for $\piz$ unambiguously distinguishes among various 
theoretical predictions for the form factors of the $\gamma^*\gamma \ra \piz$ transition. 
We have demonstrated that the non-perturbative properties 
of $\piz$ and $\etaz$ agree with each other 
which indicates that the wave functions of these two mesons are similar. 
In the $\etap$ analysis we have shown that the non-perturbative properties 
of $\etap$ differ substantially from those of $\piz$ and $\etaz$. 
Our measurement for $\etap$ provides important information for 
future theoretical investigations of the structure of this particle.

\begin{table}[h]
\begin{center}
\caption{Values of the pole-mass parameters 
$\Lambda_{\piz}$, $\Lambda_{\etaz}$ and $\Lambda_{\etap}$.}
\label{tab:final_res_pole}
\smallskip
\begin{tabular}{cc}
\hline
Decay chain & $\Lambda_{\cal R}$ (MeV) \\
\hline
$\piz\ra \gaga$ 				& 776 $\pm$ 10 $\pm$ 12 $\pm$ 16 \\
\hline
Simultaneous fit to three decay chains for $\etaz$ & 774 $\pm$ 11 $\pm$ 16 $\pm$ 22 \\
\hline
Simultaneous fit to seven decay chains for $\etap$ & 859 $\pm$ 9 $\pm$ 18 $\pm$ 20 \\
\hline
\end{tabular}
\end{center}
\end{table}

\end{document}